\documentstyle[aps,multicol,prl]{revtex}

\def\baselinestretch{1.1}

\renewcommand{\narrowtext}{\begin{multicols}{2} \global\columnwidth20.5pc}
\renewcommand{\widetext}{\end{multicols} \global\columnwidth42.5pc}
\newcommand{\be}{\begin{equation}}
\newcommand{\ee}{\end{equation}}
\newcommand{\beqs}{\begin{eqnarray}}
\newcommand{\eeqs}{\end{eqnarray}}

\newcommand{\half}{{\frac{1}{2}}}
\newcommand{\tr}{{\rm tr}}

\newcommand{\thab}{{\theta^{\alpha \beta}}}
\newcommand{\thmn}{{\theta^{\mu \nu}}}
\newcommand{\st}{{\vartheta}}
\newcommand{\ph}{{\hat \phi}}
\newcommand{\tR}{{\tilde R}}
\newcommand{\fh}{{\hat F}}

\begin{document}

\title{\Large\bf Bosonization in higher dimensions via noncommutative
field theory}
\author{ \large\bf Alexios P. Polychronakos \\[2mm] }

\address{\noindent 
Physics Department, City College of the CUNY, New York, NY 10031 \\
\rm{alexios@sci.ccny.cuny.edu} \\
 }


\maketitle

\begin{abstract}
\noindent

We propose the bosonization of a many-body fermion theory in $D$ spatial
dimensions through a noncommutative field theory on a ($2D-1$)-dimensional
space. This theory leads to a chiral current algebra over the noncommutative
space and reproduces the correct perturbative Hilbert space and excitation
energies for the fermions. The validity of the method is demonstrated by
bosonizing a two-dimensional gas of fermions in a harmonic trap.

\vspace{1mm}
\noindent
PACS-98: 04.70.Dy, 11.15.-q
\end{abstract}

\narrowtext


Bosonization of one-dimensional fermion systems is a well-established 
technique in condensed matter and field theory \cite{Bozo,WZW}.
Its usefulness lies
in the fact that the low-energy dynamics of the equivalent bosonic theory
encodes collective excitations of the fermion system and thus provides a
handle to analyzing strongly correlated fermions.

The success of the method relies on special properties
of one-dimensional spaces. Intuitively, the fermion dynamics
around the Fermi (or Dirac) sea become tantamount to 
one-dimensional wave propagation, the corresponding phonon states
encoding quantum excitations \cite{Wave}. This leads to a
mapping of states between the two systems both at the many-body
and the field theory (operator) level.

There are several approaches to extending these techniques to higher
dimensions \cite{HiBoz,PhaBoz}.
The most obvious way to proceed is by generalizing the
Fermi sea idea `pointwise' in the extra dimensions, giving rise to
the so-called radial bosonization.
This approach, however, has some obvious drawbacks that are at the
heart of the difficulties with higher dimensions. Specifically,
it produces one tower of states for each point in the extra dimensions,
thus overcounting the degrees of freedom by not taking into account the
`finite cell' structure of the sea.
Improvements to the method have been proposed, but
bosonization of higher dimensional fermion systems
remains to a large extent an open issue.

The purpose of this Letter is to propose a theory that addresses the above
problem and provides a potentially exact bosonization scheme for higher
dimensional theories. The approach is through a phase space description
of the many-body fermion problem. For $D$ spatial dimensions this leads
to a $2D-1$-dimensional nonlocal bosonic theory.
(In this respect, our approach is closest to the one in \cite{PhaBoz}.)
Only in one dimension the dimensionalities agree and
the theory becomes local. We accept this as the price of the (nontrivial)
payout of higher-dimensional bosonization.

We shall present here an outline of the new method, motivated by the
semiclassical phase space droplet picture, and demonstrate it in a
simple, solvable situation. This would make its physical
meaning and content as accessible as possible. A full derivation will be
given in a subsequent publication.

We shall consider a collection of spinless, noninteracting fermions in
$D=d+1$ spatial dimensions. (Spin and interactions can be added later without
conceptual problems.) The starting point will be the semiclassical phase
space droplet description of the system \cite{PhSp}, in the hamiltonian
formulation derived and fully exposed in \cite{AP}.
We review here the basic notions and quote the relevant results.

Particles move on a phase space of dimension $2d+2$ with coordinates
$\phi^\mu$ and a single-particle Poisson structure
\be
\{ \phi^\mu , \phi^\nu \}_{sp} = \thmn ~,~~~ \mu,\nu = 1,\dots 2d+2
\label{spPB}
\ee
For simplicity we shall assume that we have chosed Darboux coordinates
and momenta $\phi^\mu = \{ x^i , p^i \}$ so that $\thmn$ is in the
canonical form, with determinant
$\det \thmn = 1$. Particle dynamics are fixed by a single-particle
hamiltonian $H_{sp} ( \phi^\mu  )$.

The system can be described semiclassically by a state
in which each cell of volume $(2\pi\hbar)^{d+1}$ in phase space is filled by
a single fermion. This leads to a phase space `droplet' picture of
constant density $\rho_o = 1/(2\pi\hbar)^{d+1}$, determined by its
$2d+1$-dimensional boundary. We choose to parametrize the boundary in terms
of one of the phase space coordinates, say $p^1$, which becomes a function
of the remaining variables $R(\phi^\mu )$ ($\phi^\mu \neq p^1$).
It is convenient to separate in notation the variable conjugate to $p^1$,
$x^1 \equiv \sigma$ ($\{p^1 , x^1\}_{sp} = -1$),
and use middle Greek indices for the remaining $2d$
phase space variables $\phi^\alpha$ and corresponding Poisson structure
$\thab$. $\phi$ will collectively denote the coordinates $\phi^\alpha$.

The dynamics of the system are given in terms of the canonical Poisson
brackets of the field $R(\sigma,\phi )$ (not to be confused with the
single-particle brackets $\{ . , . \}_{sp}$). Using the shorthand
$R_i = R(\sigma_i , \phi_i )$, these are derived as \cite{AP}:
\be
\{ R_1 , R_2 \} = \frac{1}{\rho_o}
\left[ - \delta' (\sigma_- ) \delta (\phi_- )
- \delta (\sigma_- )
\thab \partial_\alpha R_+ \partial_\beta \delta (\phi_- ) \right]
\label{PR}
\ee
with $(\sigma_- , \phi_- )$ and $(\sigma_+ , \phi_+ )$ relative and
mid-point coordinates respectively. The hamiltonian for the field $R$ is
the sum of the single-particle hamiltonians over the bulk of the droplet
\be
H = \rho_o \int dp^1 d\sigma d^{2d} \phi H_{sp} (\sigma , \phi) \st (R-p^1 )
\label{H}
\ee
where $\st (x) = \half [1+ \rm{sgn} (x)]$ is the step function.
(\ref{PR}) and (\ref{H}) define a bosonic field theory
(in a hamiltonian setting) that describes the system semiclassically.

The above theory arlready represents an improvement over previous methods,
which essentially amounted to keeping only the first (constant) term in the
Poisson brackets. Althought at this stage it still suffers from shortcomings
similar to the ones of previous Fermi sea approaches, it will serve
as the starting point
for an exact bosonization. Below we shall simply motivate the form of the
exact theory and demonstrate its merits, leaving a full derivation and
details for a future publication.

We start by noticing that $R_+$ in the right hand side of the above Poisson
brackets can also be put to $R_1$ (or $R_2$), since the antisymmetry of
$\thab$ makes the difference irrelevant. In view of this, the right hand side
can be recast using the single-particle Poisson brackets for the
variables $\phi_1^\alpha$ as
\be
\thab \partial_\alpha R_+ \partial_\beta \delta (\phi_- ) \to
\{ R(\sigma_1 , \phi_1 ) , \delta (\phi_1 - \phi_2 ) \}_{sp1}
\ee
We now make the central claim of this paper: the exact theory will be
obtained from the above expression by `quantizing' (at the classical
level) the remaining phase space coordinates; that is, by promoting
the single-particle Poisson brackets to noncommutative Moyal brackets
$\{ . , . \}_*$ on the $2d$-dimensional phase space manifold $\phi^\alpha$:
\be
\{ R_1 , R_2 \} = \frac{1}{\rho_o}
\left[ -\delta' (\sigma_- ) \delta (\phi_- )
- \delta (\sigma_- )
\{ R_1 , \delta (\phi_- ) \}_{*1} \right]
\label{qPB}
\ee

The Moyal brackets between two functions of $\phi$ are expressed
in terms of the noncommutative Groenewald-Moyal star-product on the phase
space $\phi$ \cite{Star}:
\be
\{ F(\phi) , G(\phi) \}_* = \frac{1}{i\hbar} \left[
F(\phi) * G(\phi) - G(\phi) * F(\phi) \right]
\label{Moy}
\ee
with $\hbar$ itself being the noncommutativity parameter.
Correspondingly, the hamiltonian $H$ is given by expression (\ref{H})
but with star-products replacing the usual products between its terms.

Inherent in
the definition of the star product is a one-to-one mapping between classical
functions $F(\phi)$ and operators $\fh$ on a first-quantized phase space.
Specifically, define the set of quantum operators $\ph$ corresponding
to the residual classical phase space, satisfying
\be
[\ph^\alpha , \ph^\beta ] = i\hbar \thab ~,~~~ \alpha,\beta=1,\dots ,2d
\ee
and choose a basis $|a)$ for the (single-particle) quantum mechanical
Hilbert space on which the above operators act. This could be, e.g., the
Fock space of $d$ harmonic oscillators, or any other basis.
Classical commutative functions map to operators by adopting a
specific operator ordering $: F (\ph ) :$ in the expression
for $F(\ph)$, and star product maps to operator product:
\be
F(\phi) \leftrightarrow \fh = \, :F(\ph): ~,~~~
(F * G)(\phi) \leftrightarrow :\fh {\hat G}:
\ee
It is customary to choose the fully symmetric Weyl ordering
\be
: e^{i q^\beta \ph^\beta} : = e^{i q^\beta \ph^\beta} ~,~~~ q \in R^{2d}
\ee
in which case the corresponding star product is given in terms of the
Fourier transform of the functions as
\be
({\tilde F} * {\tilde G} )(k) = \int \frac{d^{2d} q}{(2\pi)^{2d}}
{\tilde F}(k-q) {\tilde G}(q) e^{\frac{i}{2} \hbar \thab q_\alpha k_\beta}
\label{star}
\ee
Any other appropriate ordering can be used, leading to alternative
definitions of the star product.

In the limit $\hbar \to 0$ the above becomes the usual (commutative)
convolution integral in Fourier space, while the brackets (\ref{Moy}) 
go over to standard Poisson brackets on the manifold.
For nonzero $\hbar$ the star-product and corresponding Moyal brackets
are nonlocal. The obtained theory is, therefore, nonlocal in the $2d$
dimensional phase space while it remains local in the $\sigma$ direction.

To analyze the content of the above theory, it is beneficial to perform
a change of variable in the field $R$ to a matrix representation.
We define the mapping between functions $F(\phi)$ and matrix elements
$F^{ab} = (a|\fh |b)$:
\be
F^{ab} = \int d^{2d}\phi F(\phi) C_{ab} (\phi) ~,~~
F(\phi) = \sum_{a,b} F^{ab} C_{ba} (\phi)
\ee
with $C_{ab} (\phi)$ an appropriate set of basis functions depending
on the ordering:
\be
C_{ab} (\phi) = \int \frac{d^{2d} q}{(2\pi)^{2d}}
e^{i q^\alpha \phi^\alpha} (a| : e^{-i q^\beta \ph^\beta} : |b)
\ee
In this representation, star products become matrix multiplication
and phase space integrals become traces:
\be
(F*G)^{ab} = F^{ac} G^{cb}
~,~~~
\int d^{2d} \phi F(\phi) = (2\pi\hbar)^d F^{aa}
\ee
with repeated indices summed.

The above mapping can be applied to the classical dynamical field
$R(\sigma,\phi)$ mapping it to dynamical matrix elements
$R^{ab} (\sigma)$. To express the full Poisson brackets for $R$ in 
this representation we need the matrix expression for the delta
function $\delta (\phi_1 - \phi_2 )$, with defining property
\be
\int d^{2d} \phi_1 F(\phi_1 ) \delta (\phi_1 - \phi_2 ) = F(\phi_2 )
\ee
Since $\delta (\phi_1 - \phi_2 )$ is a function of two variables, its
matrix transform in each of them will produce a symbol with four indices
$\delta^{a_1 b_1 ; a_2 b_2}$. The above defining relation in
the matrix representation becomes
\be
(2\pi\hbar)^d F^{a_1 b_1} \delta^{b_1 a_1 ; a_2 b_2} = F^{a_2 b_2}
\ee
which implies
\be
\delta^{a_1 b_1 ; a_2 b_2} = \frac{1}{(2\pi\hbar)^d} \delta^{a_1 b_2}
\delta^{a_2 b_1}
\ee

With the above, and using $\rho_o = 1/(2\pi\hbar)^{d+1}$,
the canonical Poisson brackets of the matrix $R^{ab}$
become
\beqs
\{ R_1^{ab} , R_2^{cd} \} = &-& 2\pi\hbar  \delta'
(\sigma_- ) \delta^{ad} \delta^{cb} \cr
&+& 2\pi i \delta (\sigma_- )
(R_1^{ad} \delta^{cb} - R_1^{cb} \delta^{ad} ) 
\label{alga}
\eeqs

Upon quantization of the theory, the fields $R^{ab} (\sigma)$ become operators
whose quantum commutator is given by the above Poisson brackets times $i\hbar$.
Defining, further, the Fourier modes
\be
R_k^{ab} = \int \frac{d\sigma}{2\pi\hbar} R^{ab}(\sigma) e^{-ik\sigma}
\label{Fou}
\ee
the quantum commutators become 
\be
[ R_k^{ab} , R_{k'}^{cd} ] = k \delta (k+k' ) \delta^{ad} \delta^{cb}
- R_{k+k'}^{ad} \delta^{cb} + R_{k+k'}^{cb} \delta^{ad}
\label{quantcom}
\ee
The Casimir $R_0^{aa} \equiv N$ is the total fermion number. For a compact
dimension $\sigma$, normalized to a circle of length $2\pi$, the Fourier
modes become discrete.

This is nothing but a chiral current algebra for the matrix field $R_k^{ab}$
on the unitary group of transformations of the first-quantized states $|a)$.
To make this explicit, consider for the moment that the first-quantized
Hilbert space is $K$-dimensional, that is, $a,b,c,d=1,\dots K$ (this
would be the case for a compact phase space $\{ \phi^\alpha \}$), and
define the hermitian $K \times K$ fundamental generators of
$U(K)$, $T^A$, $A=0,\dots K^2 -1$,
normalized as $\tr (T^A T^B ) = \half \delta^{AB}$
and defining the $U(M)$ structure constants $[ T^A , T^B ] = i f^{ABC} T^C$
with $f^{0AB} = 0$. Using the $T^A$ as a basis we express the quantum
commutators (\ref{quantcom}) in terms of the $R^A = \tr (T^A R)$ as
\be
[ R_k^A , R_{k'}^B ] = \half k \delta (k+k' ) \delta^{AB} +
i f^{ABC} R_{k+k'}^C
\ee
This is the so-called Kac-Moody algebra for the group $U(K)$.

The coefficient $k_{_{KM}}$ of the central extension of the Kac-Moody algebra
(the first, affine term) must be quantized to an integer to have
unitary representations. Interestingly, this coefficient in the above
commutators emerges quantized to the value $k_{_{KM}}=1$.
This is crucial for bosonization \cite{WZW}. The $k_{_{KM}}=1$ algebra has a 
unique irreducible
unitary representation over each `vacuum'; that is, over highest
weight states annihilated by all $R^A (k)$ for $k>0$ and transforming
under a fully antisymmetric $SU(K)$ representations under $T^A (0)$.
We argue that {\it these Fock-like representations correspond exactly
to the perturbative Hilbert space of excitations of the many-body
fermionic system over the full set of possible Fermi sea ground states.}
The $U(1)$ charge $R_0^0$, which is a Casimir, corresponds to the total
fermion number; diagonal operators $R_k^H $, for $k<0$ and $H$ in the
Cartan subgroup of $U(K)$ generate `radial' excitations
in the Fermi sea; while off-diagonal operators $R_k^T $, for $k<0$ and
$T$ off the Cartan subgroup, generate transitions of fermions between
different points of the Fermi sea.

Finally, the hamiltonian of the bosonic theory becomes
\be
H = \int \frac{dp^1 d\sigma}{2\pi\hbar} \tr H_{sp} (\sigma , p^1, {\hat \phi})
\st (R-p^1 )
\label{HQ}
\ee
where $p^1$ remains a scalar integration parameter while $\hat \phi$
and $R$ are matrices as before.
Clearly there are issues of ordering in the above expression, both
quantum and matrix, just as in standard $1+1$-dimensional bosonization.

It is clear that for $D=1$ our theory reproduces standard
bosonization. In this case there is no residual phase space and we recognize
$R(\sigma)$ as the derivative of the bosonized field $\partial_\sigma \Phi$ representing
the chiral density in one dimension. (Two independent chiral fields $R_\pm$,
representing the two intercepts of the compact droplet boundary with the
line $x$=constant, are needed.) The expression of the hamiltonian
(\ref{HQ}) for $H_{sp} = p$ or $H_{sp} = p^2 /2$ assumes the usual
relativistic or nonrelativistic bosonized form $H = (\partial_x \Phi )^2 /2$
or $H = (\partial_x \Phi)^3 / 6$, respectively, fully recovering the
standard results.

To demonstrate the applicability of this theory we shall work out
explicitly the simplest nontrivial example of higher-dimensional
bosonization: a system of $N$ noninteracting two-dimensional fermions
in a harmonic oscillator potential. The single-particle
hamiltonian is
\be
H_{sp} = \half (p_1^2 + x_1^2 + p_2^2 + x_2^2 )
\ee
where for simplicity we chose the oscillator to be isotropic
and of unit frequency. The single-body spectrum is the direct sum of two
simple harmonic oscillator spectra, $E_{mn} = \hbar (m+n+1)$, $m,n=0,1,\dots$.
Calling $m+n+1=K$, the energy levels are $E_K = \hbar K$ with degeneracy $K$.

The $N$-body ground state consists of fermions filling states $E_K$
up to a Fermi level $E_F = \hbar K_F$. In general, this state is degenerate,
since the last energy level of degeneracy $K_F$ is not fully occupied.
Specifically, for a number of fermions $N$ satisfying
\be
N = \frac{K_F (K_F -1)}{2} + M ~,~~~ 0 \le M \le K_F
\ee
the Fermi sea consists of a fully filled bulk (the first term above)
and $M$ fermions on the $K_F$-degenerate level at the surface.
The degeneracy of this many-body state is 
\be
g (K_F , M) = \frac{K_F !}{M! (K_F -M)!}
\ee
representing the ways to distribute the $M$ last fermions over $K_F$ states,
and its energy is
\be
E (K_F , M) = \hbar \frac{K_F (K_F -1) (2K_F -1)}{6} + \hbar K_F M
\ee
Clearly the vacua $(K_F , M=K_F )$ and $(K_F +1 , M=0)$ are identical.
Excitations over the Fermi sea come with energies in integer multiples of
$\hbar$ and degeneracies according to the possible fermion arrangements. 

For the bosonized system we choose polar phase space variables,
\be
h_i = \half (p_i^2 + x_i^2 )~,~~~ 
\theta_i = \arctan \frac{x_i}{p_i} ~,~~i=1,2
\ee
in terms of which the single-particle hamiltonian and Poisson structure is
\be
\{ \theta_i , h_j \}_{sp} = \delta_{ij} ~,~~~ H_{sp} = h_1 + h_2
\ee
For the droplet description we take $h_1 = R$ and $\theta_1 = \sigma$
which leaves $\{ h_2 , \theta_2 \} \sim \{ x_2 , p_2 \}$
as the residual phase space. The bosonic hamiltonian is
\be
H = \frac{1}{(2\pi\hbar)^2} \int d\sigma dh_2 d\theta_2 (\half R^2 + h_2 R )
\label{Harm}
\ee
The ground state is a configuration with $R + h_2 = E_F$. The nonperturbative
constraints $R>0$, $h_2 >0$ mean that the range of $h_2$ is $0<h_2 <E_F$.

We `quantize' the single-particle residual phase space $(h_2 , \theta_2 )$
by defining oscillator states $|a)$, $a=0,1,2,\dots$
satisfying ${\hat h}_2 |a) = \hbar (a+\half) |a)$. The nonperturbative
constraint for $h_2$ is implemented by restricting to the $K_F$-dimensional
Hilbert space spanned by $a=0,1,\dots K_F$ with $E_F = \hbar K_F -1$.
In the matrix representation $R^{ab}$ becomes a $U(K_F )$ current algebra.
We also Fourier transform in $\sigma$ as in (\ref{Fou}) into discrete modes
$R_n^{ab}$, $n = 0, \pm 1 , \dots$ ($\sigma$ has a period $2\pi$).
The hamiltonian (\ref{Harm})
has no matrix ordering ambiguities (being quadratic in $R$ and $h_2$)
but it needs quantum ordering. Just as in the $1+1$-dimensional case,
we normal order by pulling negative modes $N$ to the left
of positive ones. The result is
\be
\frac{H}{\hbar} = \sum_{n<0} R_{-n}^{ab} R_n^{ba} 
+ \half R_0^{ab} R_0^{ba} + (a+\half) R_0^{aa} 
\label{Hq}
\ee

To analyze the spectrum of (\ref{Hq}) we perform the change of variables
\be
\tR_n^{ab} = R_{n-a+b}^{ab} + (a-K_F +1 ) \delta^{ab} \delta_n
\ee
The new fields $\tR$ satisfy the same Kac-Moody algebra as $R$. The
hamiltonian (\ref{Hq}) becomes
\beqs
\frac{H}{\hbar} = \sum_{n<0} && \tR_{-n}^{ab} \tR_n^{ba} 
+ \half \tR_0^{ab} \tR_0^{ba} + (K_F - \half ) \tR_0^{aa} \cr
&&+ \frac{K_F (K_F -1)(2K_F -1)}{6} 
\label{Htq}
\eeqs
The above is the standard quadratic form in $\tR$ plus a constant and
a term proportional to the $U(1)$ charge $\tR_0^{aa} = N - K_F (K_F -1)/2 $.

The ground state consists of the vacuum multiplet $|K_F, M\rangle$,
annihilated by all positive modes $\tR_n$ and transforming in the $M$-fold
fully antisymmetric irrep of $SU(K_F )$ ($0 \le M \le K_F -1$), with
degeneracy equal to the dimension of this representation $K_F ! /M! 
( K_F -M )!$. The $U(1)$ charge of $\tR$ is given by the number of boxes
in the Young tableau of the irreps, so it is $M$. The fermion
number is, then, $N = K_F (K_F -1)/2 +M$. Overall, we have a full
correspondence with the many-body fermion ground states found before;
the state $M = K_F$ is absent, consistently with the fact that the
corresponding many-body state is the state $M=0$ for a shifted $K_F$.

The energy of the above states consists of a constant plus a dynamical
contribution from the zero mode $\tR_0$.
The quadratic part contributes $\half \hbar M$, while the linear part
contributes $\hbar (K_F - \half) M$. Overall, the energy is
$\hbar K_F (K_F -1 ) (2K_F -1)/6 + \hbar K_F M$, also in agreement with
the many-body result.

Excited states are obtained by acting with creation operators $\tR_{-n}$
on the vacuum, introducing $n$ quanta of energy. Due to the presence
of zero-norm states, the corresponding Fock representation truncates in
just the right way to reproduce the states of second-quantized fermions
with an $SU(K_F )$ internal symmetry and fixed total fermion number.
These are, again, into one-to-one correspondence with
the excitation states of the many-body system, built as towers of 
one-dimensional excited Fermi seas over single-particle states $E_{m,n}$, 
one tower for each value of $n$, with the correct excitation energy.
We have the nonperturbative constraint
$0 \le n < K_F$, as well as constraints related
to the non-depletion of the Fermi sea for each value of $n$, just as
in the one-dimensional case. The off-diagonal operators $\tR_n^{ab}$ create
particle transitions between towers, with the total particle number fixed to $N$
by the value of the $\tR_0$.

We conclude by stating that the proposed bosonized theory is defined in a
hamiltonian setting. The transition to a lagrangian
requires `inverting' the canonical Poisson brackets to a two-form
$\omega$ and corresponding connection $\omega = dA$.
The resulting action will contain the noncommutative generalization of the
Wess-Zumino term, expressed in terms of a star-unitary field $U$
given by the star-exponential of a real field $U = \exp_* (i \Phi)$
\cite{NCWZ}.
This will be presented in a future publication.

\def\baselinestretch{1.0}

\widetext

\end{document}